\documentclass[aps,amsmath,a4paper,prb,twocolumn]{revtex4}

\usepackage{graphicx}
\usepackage[dvips]{color}

\newcommand {\be} {\begin{equation}}
\newcommand {\ee} {\end{equation}}
\newcommand {\bea} {\begin{eqnarray}}
\newcommand {\eea} {\end{eqnarray}}
\newcommand {\bes} {\begin{displaymath}}
\newcommand {\ees} {\end{displaymath}}
\newcommand {\beas} {\begin{eqnarray*}}
\newcommand {\eeas} {\end{eqnarray*}}

\begin{document}

\title{Single-file dynamics with different diffusion constants}

\author{Tobias Ambj\"ornsson}
\email{ambjorn@mit.edu}
\affiliation{Department of Chemistry, Massachusetts
Institute of Technology, Cambridge, MA 02139}

\author{Ludvig Lizana}
\affiliation{The Niels Bohr Institute, Blegdamsvej 17, 2100 Copenhagen, 
Denmark.}

\author{Michael A. Lomholt}
\affiliation{MEMPHYS - Center for Biomembrane Physics,
Department of Physics and Chemistry,
University of Southern Denmark,
Campusvej 55, 5230 Odense M, Denmark}

\author{Robert J. Silbey}
\affiliation{Department of Chemistry, Massachusetts 
Institute of Technology, Cambridge, MA 02139.}

\date{\today}



\begin{abstract}
We investigate the single-file dynamics of a tagged particle in a
system consisting of $N$ hardcore interacting particles (the particles
cannot pass each other) which are diffusing in a one-dimensional
system where the particles have {\em different} diffusion
constants. For the two particle case an exact result for the
conditional probability density function (PDF) is obtained for
arbitrary initial particle positions and all times. The two-particle
PDF is used to obtain the tagged particle PDF. For the general
$N$-particle case ($N$ large) we perform stochastic simulations using
our new computationally efficient stochastic simulation technique
based on the Gillespie algorithm. We find that the mean square
displacement for a tagged particle scales as the square root of time
(as for identical particles) for long times, with a prefactor which
depends on the diffusion constants for the particles; these results
are in excellent agreement with very recent analytic predictions in
the mathematics literature.

\end{abstract}

\maketitle


\section{Introduction} \label{sec:Introduction}

Crowding effects are ubiquitous in cells \cite{Luby-Phelps} - large
macromolecules in cells reduce the diffusion rates of particles,
influence the rates of biochemical reactions and bias the formation of
protein aggregates \cite{Ellis}. Furthermore, devices used in
nanofluidics are becoming smaller; crowding and interactions
effects between particles are therefore of increasing importance also in this
field.

An example of a system where crowding is dominant is the diffusion of
hardcore interacting particles (the particles cannot pass each other)
in one dimension, so called single-file diffusion. For single-filing
systems the particle order is conserved over time $(t)$ resulting in
interesting dynamical behavior for a tagged particle, quite different
from that of classical diffusion.  Examples found in nature are ion or
water transport through pores in biological membranes~\cite{HOKE},
one-dimensional hopping conductivity~\cite{MR} and channeling in
zeolites~\cite{KKDGPRSUK}. Furthermore, in biology there are examples
where the fact that particles cannot overtake one another are of
importance: for instance, DNA binding proteins diffusing along a DNA
chain \cite{Berg,Halford,Lomholt}.  Single-file diffusion has also
been observed in a number of experiments such as in colloidal systems
and ring-like constructions.~\cite{CJG,LKB,WBL} One of the most
apparent characteristics of single-file diffusion is that the mean
square displacement (MSD) $\langle (x_{\cal T}-x_{{\cal T},0}
)^2\rangle $ (the brackets denote an average over thermal noise and
initial positions of non-tagged particles, $x_{\cal T}$ is the tagged
particle position and $x_{{\cal T},0}$ is the initial position of the
tagged particle) of a tagged particle is proportional to $t^{1/2}$ for
long times in an infinite system with a fixed particle concentration;
the corresponding probability density function (PDF) of the tagged
particle position is Gaussian. The first study showing the $t^{1/2}$
behavior of the MSD and the fact that the PDF is Gaussian is found in
Ref. \onlinecite{HA}. Subsequent studies include
Refs. \onlinecite{LE,Beijeren_83,Hahn_95,Arratia_83,CA,Taloni_06}.
The $t^{1/2}$-law and Gaussian behavior for long times has proven to
be of general validity for identical strongly overdamped particles
where mutual passage of the particles is excluded, for arbitrary
short-range interactions between particles. \cite{Kollmann_03}
Recently, a generalized central limit theorem was proved for the
tagged particle motion.  \cite{Jara_Landim_06} It is interesting to
note that a mean square fluctuation that scales as $t^{1/2}$ also
occurs for monomer dynamics in a polymer within the Rouse model.
\cite{Khokhlov,Shusterman_04} We point out that anomalous scaling of
the MSD with time, i.e., that $\langle (x_{\cal T}-x_{{\cal T},0}
)^2\rangle $ is {\em not} proportional to $t$, can occur also due to
long waiting times between particle jump events (when the waiting time
distribution has a divergent first moment). \cite{MEKL1,MEKL2}
However, for such processes the PDF is not Gaussian; the anomalous
behavior in single-file systems is {\em not} due to long waiting time
densities but rather due to strong correlations between particles.

\begin{figure}
 \includegraphics[width=8cm]{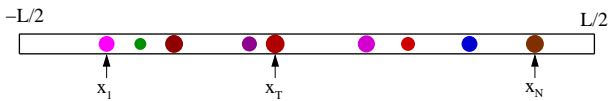}
 \caption{Cartoon of the problem considered in this study: $N$
   particles are diffusing in a one-dimensional system. 
   Particle $j$ ($j=1,...,N$) has coordinate $x_j$, initial
   coordinate $x_{j,0}$ and diffusion constant $D_j$. The particles
   cannot overtake, hence at all times we have $x_j<x_{j+1}$. In our
   analytic calculation for the two-particle case the system size is
   assumed to be infinite. In the stochastic simulations we assume a
   system of finite length $L$, with reflecting boundary conditions at
   $x=\pm L/2$.}
 \label{fig:cartoon}
\end{figure}

Although much work has been dedicated to single-file diffusion of
identical particles, fewer studies has addressed the problem of
diffusion of hardcore particles with different diffusion
constants. This type of system could be of interest, for instance, for
protein diffusion along a DNA chain (there is a plethora of DNA
binding proteins).  The single-file system with different diffusion
constants is illustrated in Fig. \ref{fig:cartoon}: The particles each
have coordinates $\vec{x}=(x_1,x_2,...,x_N)$ and initial coordinates
$\vec{x}_0=(x_{1,0},x_{2,0},...,x_{N,0})$. Due to the hardcore
interaction the particles cannot pass each other, and therefore retain
their order at all times, i.e.,
  \be\label{eq:region_Rp1_N}
 -\frac{L}{2}<x_1<x_2<...<x_N<\frac{L}{2}. 
  \ee
where $L$ is the length of the system (and we assumed the ends of the
system, at $\pm L/2$, to be reflecting).  Particle $j$ has diffusion
constant $D_j$ ($j=1,..,N$). The spatial distribution of the particles
as a function of time is contained in the $N$-particle conditional PDF
${\cal P}(\vec{x},t|\vec{x}_0)$; the equations for this quantity were
given in Ref. \onlinecite{Lizana_Ambjornsson} (with obvious modifications to
account for the different diffusion constants). We are particularly
interested in the dynamics of a tagged particle with coordinate
$x_{\cal T}$ with initial position $x_{{\cal T},0}$, which
mathematically is obtained by integrating ${\cal
P}(\vec{x},t|\vec{x}_0)$ over all coordinates and initial positions
except for $x_{\cal T}$ and $x_{{\cal T},0}$.
\cite{RKH,Lizana_Ambjornsson}

To our knowledge, the only studies investigating the type of
single-file system described above are
Refs. \onlinecite{Aslangul_00}, \onlinecite{Brzank2} and
\onlinecite{Jara_Gonzalves}. In Ref. \onlinecite{Aslangul_00} the particles
were assumed to be initially placed at the {\em same} position. Also,
the 'annealed' case, where the diffusion constants were randomized
between the particles for each new ensemble, was considered. In Ref.
\onlinecite{Brzank2} the hydrodynamic behavior of a two-component (two
different kinds of particles) single-file system with boundary
injection and extraction were considered. Very recently in the
mathematics literature, the asymptotic behavior for long times of a
tagged particle in a single-file system with different diffusion
constants was obtained for the 'quenched' case (i.e., the diffusion
constants are the same for each ensemble) for hopping dynamics on a
lattice. \cite{Jara_Gonzalves}

In this study we extend the results from previous studies by (i)
analytically solving the problem of diffusion of two hardcore
interacting particle with different diffusion constants in which the
initial positions for the two particles are {\em arbitrary}, and valid
for {\em all} times. The study of diffusion with arbitrary initial
conditions is important in the field of single-file diffusion since in
the derivation of the $t^{1/2}$-law it is assumed that the particles
are initially randomly distributed. (ii) We introduce a new fast
stochastic scheme tailored for interacting particle systems with
different diffusion constants. (iii) For the general $N$-particle case
we verify the asymptotic results obtained in
Ref. \onlinecite{Jara_Gonzalves} for long times and large $N$ using
our new stochastic algorithm, and illustrate the behavior for shorter
times.

\section{Two different hardcore interacting particles}\label{sec:ProblemDef}

We consider a system with two hardcore interacting particles diffusing
in an infinite one-dimensional system, see Fig.~\ref{fig:int_area} (top).

\subsection{Equations of motion}

\begin{figure}
\includegraphics[width=8cm]{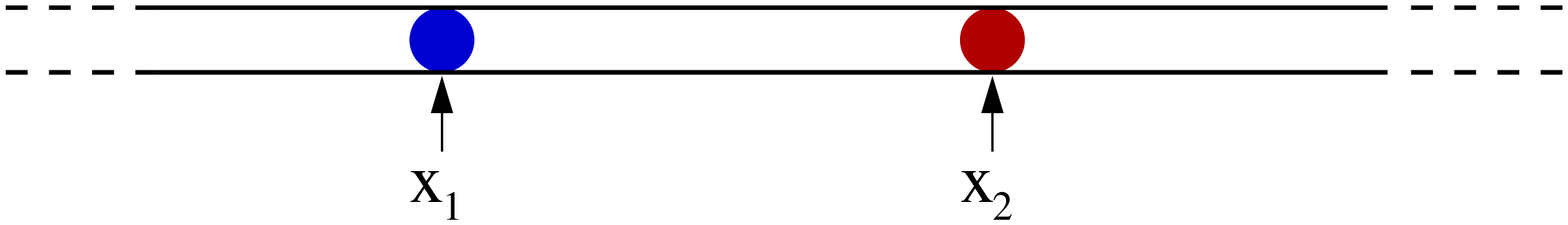}\newline\\ 
\includegraphics[width=6cm]{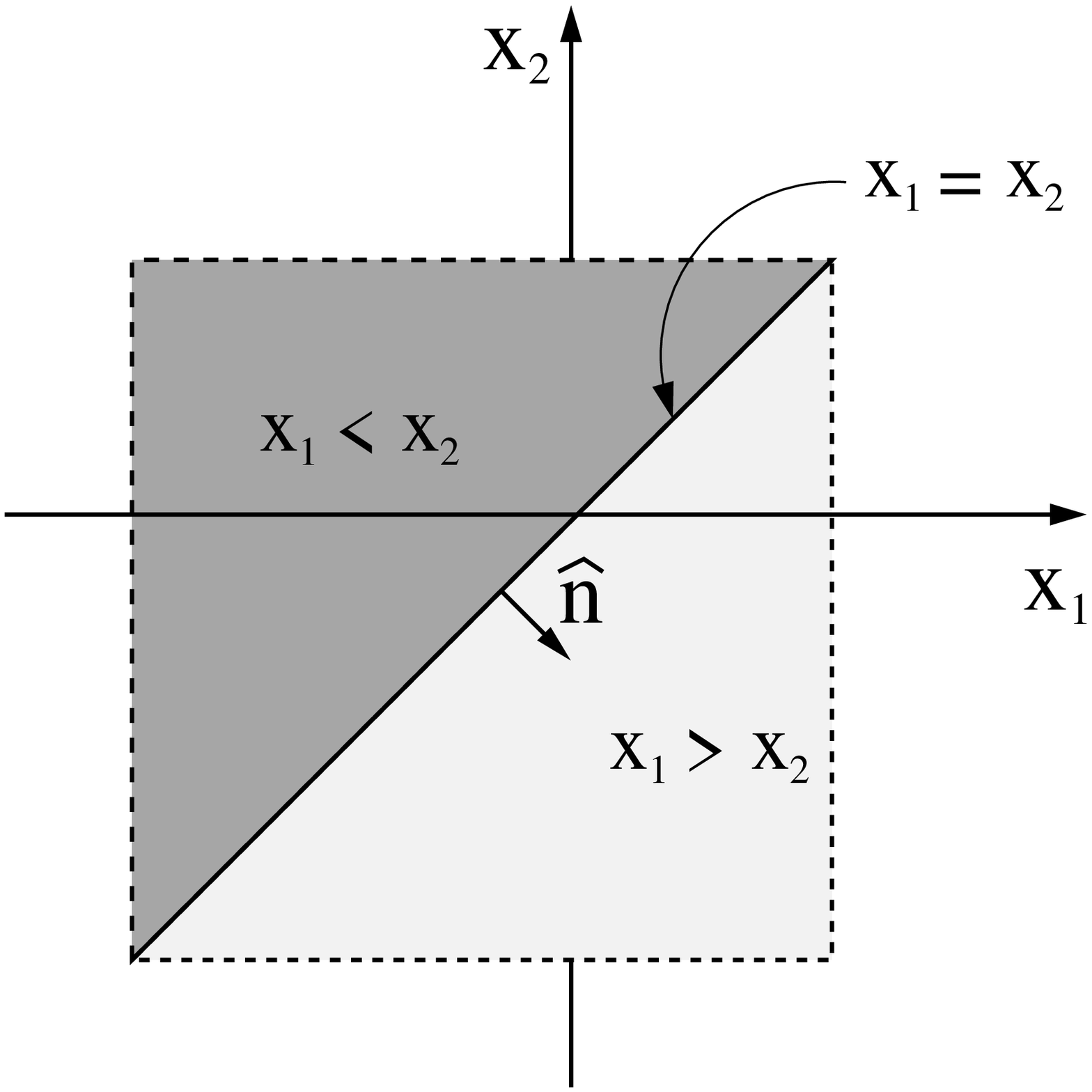}
 \caption{ (top) Cartoon of the problem considered in Sec.
   \ref{sec:ProblemDef}: two particles are diffusing in a
   one-dimensional system. Particle $j$ ($j=1,2$) has coordinate
   $x_j$, initial coordinate $x_{j,0}$ and diffusion constant $D_j$; in
   general $D_1\neq D_2$. The particles cannot overtake, hence at all
   times we have $x_1<x_2$. In our analytic calculation the system
   size is assumed to be infinite. In the stochastic simulations we
   assume a system of finite length $L$, with reflecting boundary
   conditions at $\pm L/2$.  (bottom) Phase space region ${\cal R}$
   (the darker upper area, $x_1<x_2$) for two hardcore interacting
   particles. For the analytic solutions ${\cal R}$
   extends to $\pm \infty$.}
 \label{fig:int_area}
\end{figure}

The particles each
have coordinates $\vec{x}=(x_1,x_2)$ and initial
coordinates $\vec{x}_0=(x_{1,0},x_{2,0})$. The hardcore
interaction prevents the particles from passing each other:
  \be\label{eq:region_Rp1}
 -\infty<x_1<x_2<\infty. 
  \ee
We denote the phase-space region spanned by coordinates $\vec{x}$
satisfying Eqs. (\ref{eq:region_Rp1}) by ${\cal R}$, see
Fig. \ref{fig:int_area} (bottom).
The temporal behavior of the spatial distribution of the particles is 
contained in the PDF ${\cal P}(\vec{x},t|\vec{x}_0)$ which is governed by
\be\label{eq:DiffEq}
 \frac{ \partial {\cal P}(\vec{x},t|\vec{x}_0)}{\partial t} = 
   \left(
    D_1 \frac{\partial^2}{\partial x_1^2}+D_2\frac{\partial^2}{\partial x_2^2}
  \right)
  {\cal P}(\vec{x},t|\vec{x}_0),
\ee
for $\vec{x}\in {\cal R}$ [${\cal P}(\vec{x},t|\vec{x}_0)\equiv 0$ outside
${\cal R}$] and $D_1$ ($D_2$) is the diffusion constant for particle 1
(particle 2). The initial condition is
\be\label{eq:InitCond}
 {\cal P}(\vec{x},0|\vec{x}_0) = \delta(x_1-x_{1,0})\delta(x_2-x_{2,0})
\ee
where $\delta(z)$ is the Dirac delta-function. 
The fact that the particles cannot pass each other is described by
\be\label{eq:NopassCond}
      (D_1 \frac{\partial}{\partial x_1}- D_2 \frac{\partial}{\partial x_2})
   {\cal P}(\vec{x},t|\vec{x}_0)|_{x_1=x_2} = 0.
\ee
The above relation is a no flux condition for the normal component of
the flux vector across the line $x_1=x_2$, see Fig.
\ref{fig:int_area} (bottom): Eq. (\ref{eq:DiffEq}) can be written as a
continuity equation $\partial {\cal P}/\partial t =-\vec{\nabla} \cdot
\vec{{\cal J}}$, where the flux vector is $\vec{{\cal J}}=-(\hat{x}_1
D_1 \partial {\cal P}/\partial x_1 + \hat{x}_2 D_2\partial {\cal
P}/\partial x_2)$, and $\hat{x}_1$ ($\hat{x}_2$) is a unit vector in
the $x_1$ ($x_2$) direction. The outward normal to the $x_1=x_2$
interface is (see Fig. \ref{fig:int_area})
$\hat{n}=(\hat{x}_1-\hat{x}_2)/\sqrt{2}$ which allows us to write
Eq. (\ref{eq:NopassCond}) as $\hat{n}\cdot \vec{{\cal
J}}|_{x_1=x_2}=0$. This reflecting condition guarantees that the
probability in the allowed phase-space region ${\cal R}$ is conserved
at all times as it should.

\subsection{Solution for two-particle PDF}

In order to solve the equations specified in the previous subsection we 
make the variable transformation:
  \bea
X&=&\frac{1}{2}\left(  \sqrt{\frac{D_2}{D_1}}x_1 + \sqrt{\frac{D_1}{D_2}} x_2\right)\nonumber\\
q&=&x_2-x_1.
  \eea
Eqs. (\ref{eq:DiffEq}), (\ref{eq:InitCond}) and (\ref{eq:NopassCond})
then become
  \bea
\label{eq:P^X^q_eq}
\frac{\partial {\cal P}(X,q,t)}{\partial t}&=&\left(D^X \frac{\partial^2}{\partial X^2}+D^q \frac{\partial^2}{\partial q^2}\right) {\cal P}(X,q,t)\nonumber\\
\left.\frac{\partial {\cal P}(X,q,t)}{\partial q}\right|_{q=0}&=&0\nonumber\\
{\cal P}(X,q,t\rightarrow 0)&=&\gamma \delta(X-X_0)\delta(q-q_0)
  \eea
where $\gamma=(D_1+D_2)/(2\sqrt{D_1D_2})$, $X_0=\sqrt{D_2/D_1}x_{1,0} +
\sqrt{D_1/D_2} x_{2,0}$ and $q_0=x_{2,0}-x_{1,0}$ and we introduced the
effective diffusion constants
  \bea
D^X&=&\frac{D_1+D_2}{4}\nonumber\\
D^q&=&D_1+D_2
  \eea
For the case of identical diffusion constants,
$D_1=D_2=D$ the equations above express the fact
that the relative coordinate $q$ diffuses with a diffusion constant
$2D$, whereas the center-of-mass coordinate $X$ diffuses with a
diffusion constant $D/2$. \cite{Aslangul_99}

Eq. (\ref{eq:P^X^q_eq}) allows a product solution of the form
  \be
{\cal P}(X,q,t)={\cal P}^X(X,t){\cal P}^q(q,t)\label{eq:P^X_s}
  \ee
where 
  \be
{\cal P}^X(X,t)=\frac{\gamma}{(4\pi D^X t)^{1/2}}\exp(-\frac{(X-X_0)^2}{4D^Xt})\label{eq:P^X}
  \ee
and the solution for ${\cal P}^q(q,t)$ is obtained via the method of images
\cite{RKH} according to
  \bea
{\cal P}^q(q,t)&=&\theta(q)\frac{1}{(4\pi D^q t)^{1/2}}\nonumber\\
&&\hspace{-1cm}\times \left( \exp(-\frac{(q-q_0)^2}{4D^qt})+
\exp(-\frac{(q+q_0)^2}{4D^qt}) \right)\label{eq:P^q}
  \eea
where $\theta(q)$ is the Heaviside step function, $\theta(q>0)=1$ and
$\theta(q<0)=0$. Returning to our original coordinates,
Eq. (\ref{eq:P^X_s}), (\ref{eq:P^X}) and (\ref{eq:P^q}) become,
after some algebraic manipulations:
  \bea
\label{eq:P_x1_x2}
{\cal P}(\vec{x},t|\vec{x}_0)&=&\theta(x_2-x_1) \frac{1}{(4\pi D_1 t)^{1/2}}\frac{1}{(4\pi D_2 t)^{1/2}} \nonumber\\
&&\hspace{-1cm}\times [ \exp(-\frac{(x_1-x_{1,0})^2}{4D_1 t})\exp(-\frac{(x_2-x_{2,0})^2}{4D_2t})\nonumber\\
&&\hspace{-1cm} + \exp(-\frac{(x_1-x^i_{1,0})^2}{4D_1 t})\exp(-\frac{(x_2-x^i_{2,0})^2}{4D_2t}) ]
  \eea
where the effective image initial positions are
  \bea\label{eq:x^i}
x^i_{1,0}&=&\frac{D_2-D_1}{D_1+D_2}x_{1,0}+\frac{2D_1}{D_1+D_2} x_{2,0}\nonumber\\
x^i_{2,0}&=&\frac{2D_2}{D_1+D_2}x_{1,0}+\frac{D_1-D_2}{D_1+D_2} x_{2,0}
  \eea
Notice that we have $x^i_{2,0}-x^i_{1,0}=-(x_{2,0}-x_{1,0})$, i.e. the
distance between the image initial positions is the same as the
distance between the initial positions. Eq. (\ref{eq:x^i}) is a
non-trivial extension of the image positions for identical particles
or for a system where the particles initially start out at the same
point in space: For $D_1=D_2$ we have $x^i_{1,0}=x_{2,0}$ and
$x^i_{2,0}=x_{1,0}$ as it should. \cite{Fisher_84}  For the case
$x_{1,0}=x_{2,0}=0$ the results above reduce to the results obtained
in Ref. \onlinecite{Aslangul_00}. Note that, in contrast, when
$D_1\neq D_2$ and $x_{1,0}\neq x_{2,0}$ the image initial positions,
Eq. (\ref{eq:x^i}), depend on $D_1$ and $D_2$.

It is interesting to compare the above result for ${\cal
P}(\vec{x},t|\vec{x}_0)$ to that of a Bethe-ansatz
\cite{SC,Batchelor_07}. It is straightforward to show that Eq.
(\ref{eq:P_x1_x2}) can be written:
  \bea\label{eq:P_x1_x2_Bethe}
{\cal P}(\vec{x},t|\vec{x}_0)&=&\theta (x_2-x_1)\int_{-\infty}^\infty \frac{dk_1}{2\pi} \int_{-\infty}^\infty \frac{dk_2}{2\pi}\nonumber\\
&&\hspace{-1.5cm} e^{-D_1k_1^2 t}e^{-D_2k_2^2t}e^{-ik_1x_{1,0}}e^{-ik_2 x_{2,0}} \nonumber\\
&&\hspace{-2cm} \times [e^{ik_1x_1}e^{ik_2 x_2} +g(k_1,k_2,x_1,x_2)e^{ik_2 x_1}e^{ik_1 x_2 }] 
  \eea
where
  \be
g(k_1,k_2,x_1,x_2)=\exp [ \frac{D_1-D_2}{D_1+D_2}(k_1+k_2)(x_2-x_1) ]
  \ee
We note that Eq. (\ref{eq:P_x1_x2_Bethe}) has the form of a Bethe
ansatz \cite{SC}, where the ``scattering coefficient'' $g$ depends
on $x_1$ and $x_2$ (in the standard Bethe ansatz the scattering
coefficient only depends on $k_1$ and $k_2$); the standard
Bethe-ansatz satisfies the equations of motion and the boundary
conditions for {\em fixed} $k_1$ and $k_2$; in contrast, the solution
above does not - it is only after the integrations over $k_1$ and $k_2$
are performed [with the appropriate $x_1$ and $x_2$ dependent
``mixing'' of $k_1$ and $k_2$ from the 2nd term in Eq.
(\ref{eq:P_x1_x2_Bethe})] that the correct solution for ${\cal
P}(\vec{x},t|\vec{x}_0)$ is obtained. For the case of identical diffusion
constants $D_1=D_2=D$ the mixing of $k_1$ and $k_2$ in $g$ is absent
and we have $g=1$ in agreement with previous studies.
\cite{Lizana_Ambjornsson}


\subsection{Tagged particle PDF}\label{sec:P_bar}

By integrating the two-particle PDF we obtain the tagged particle
PDF. The tagged PDF (for fixed initial
positions) for particle 1 is
$\rho_1(x_1,t|\vec{x}_0)=\int_{x_1}^\infty dx_2 {\cal
P}(\vec{x},t|\vec{x}_0)$. Explicitly, using Eq.
(\ref{eq:P_x1_x2}), we have:
  \bea\label{eq:P1}
\rho_1(x_1,t|\vec{x}_0)&=&\frac{1}{(4\pi D_1 t)^{1/2}} \exp( - \frac{(x_1-x_{1,0})^2}{4D_1 t}) \nonumber\\
&& \times \frac{1}{2}{\rm erfc} (\frac{x_1-x_{2,0}}{\sqrt{4D_2t}}) \nonumber\\
&& +\frac{1}{(4\pi D_1 t)^{1/2}} \exp( - \frac{(x_1-x^i_{1,0})^2}{4D_1 t}) \nonumber\\
&&\times \frac{1}{2}{\rm erfc} (\frac{x_1-x^i_{2,0}}{\sqrt{4D_2t}})
  \eea
where ${\rm erfc}(z)=1-{\rm erf}(z)$ is the complementary error
function, with ${\rm erf}(z)=(2/\sqrt{\pi}) \int_0^z dy \exp(-y^2)$
being the error function \cite{ABST} and $x_{1,0}^i$ and $x_{2,0}^i$
are given in Eq. (\ref{eq:x^i}). The tagged particle PDF for particle
2 $\rho_2(x_2,t|\vec{x}_0)=\int_{-\infty}^{x_2}dx_1 {\cal
P}(\vec{x},t|\vec{x}_0)$ is obtained by the replacements
$x_1\leftrightarrow -x_2$, $x_{1,0}\leftrightarrow -x_{2,0}$,
$x^i_{1,0}\leftrightarrow -x^i_{2,0}$, and $D_1\leftrightarrow D_2$ in
Eq. (\ref{eq:P1}).  We point out that Eq. (\ref{eq:P1}) does {\em not}
give an MSD which scales with time as $t^{1/2}$ (see Introduction); it
is only in the limit of a large number of hardcore interacting
particles that $\rho_j(x_j,t|\vec{x}_0)$ (with an additional average
over the initial position of non-tagged particles), for a center
particle, becomes a Gaussian with a width that scales as the square
root of time, see next section. 

A limit not accessible through
previous approaches \cite{Fisher_84,Aslangul_00} is that where one of
the particles is immobile $D_2=0$: setting $x_{20}=0$ for convenience
and taking the limit $D_2\rightarrow 0$ Eq. (\ref{eq:P1}) becomes
  \bea
\rho_1(x_1,t|\vec{x}_0) |_{D_2=0}&=& \theta(-x_1) \frac{1}{(4\pi
D_1 t)^{1/2}}\nonumber\\
&&\hspace{-3cm}\times[ \exp( - \frac{(x_1-x_{1,0})^2}{4D_1 t}) + \exp( -
\frac{(x_1+x_{1,0})^2}{4D_1 t})],
  \eea
in agreement with the diffusion of a particle near a reflecting wall
as it should. We have above used the fact that ${\rm erf} (\pm
\infty)=\pm 1$.

\begin{figure}
 \includegraphics[width=8cm]{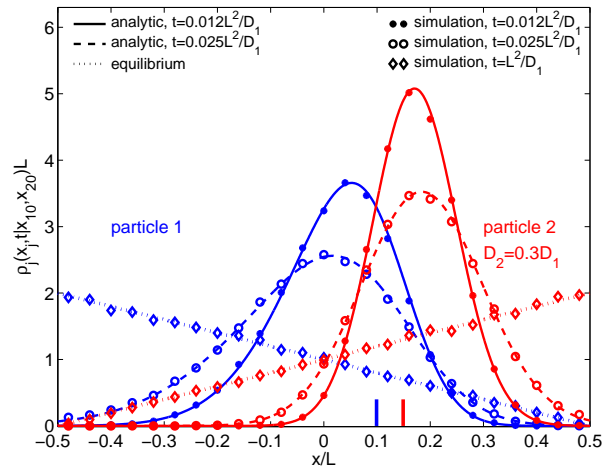}
 \caption{Tagged particle PDF $\rho_j(x_j,t|\vec{x}_0)$ for two
hardcore interacting particles ($j=1,2$). The solid and dashed
leftmost (blue) [rightmost (red)] curves corresponds to
the tagged particle PDF for particle 1 [particle 2] at different times
as given in Eq. (\ref{eq:P1}).  The ratio of the diffusion constants
is $D_2/D_1=0.3$. The symbols correspond to the results of the
Gillespie simulation, ensemble-averaged over $n_{\rm ens}=50000$
ensembles for $M=500$ lattice sites; the data was binned into 25
bins. The vertical bars at the bottom of the figure corresponds to the
initial positions for the two particles ($x_{1,0}=0.10L$ and
$x_{2,0}=0.15L$). For long time $t>L^2/D$ the equilibrium is reached -
the dashed lines correspond to the analytic result as given in
Eq. (\ref{eq:P_eq}). }
 \label{fig:pdf}
\end{figure}

In Fig. \ref{fig:pdf} we illustrate the results for the tagged
particle PDFs as given in Eq. (\ref{eq:P1}). We compare to stochastic
simulations using a new stochastic algorithm described in Appendix
\ref{sec:Gillespie}. In the simulation we assume a finite box; the
main effect of the finite box (with reflecting conditions) is to
modify the long-time limit (i.e. for $t\gg L^2/D_j$), to the
following equilibrium PDFs:
\cite{Lizana_Ambjornsson} 
 \bea\label{eq:P_eq}
 \rho^{\rm eq}(x_1) &=&  \frac{2}{L^2} \left(\frac{L}{2}-x_1\right),
\nonumber\\
\rho^{\rm eq}(x_2) &=& \frac{2}{L^2} \left(\frac{L}{2}+x_2\right)
  \eea
The results given in Eq. (\ref{eq:P_eq}) is obtained by direct
integration of the two-particle equilibrium PDF $P^{\rm
eq}(x_1,x_2)=2\theta(x_2-x_1)/L^2$. The equilibrium results are
independent on $D_1$ and $D_2$ as it should.  In Fig. \ref{fig:pdf} we
illustrate the result of a Gillespie simulation using $n_{\rm
ens}=50000$ ensembles on a lattice with $M=500$ lattice points, and
compare to the PDF Eq. (\ref{eq:P1}) as well as the equilibrium PDF,
Eq. (\ref{eq:P_eq}). We notice excellent agreement within the limit
of applicability.

It remains a challenge to generalize the results in this section to
the $N$-particle case and arbitrary times. In the next section we
perform stochastic simulations for $N$ particles and verify the
asymptotic results in Ref. \onlinecite{Jara_Gonzalves} for long times
and $N$ large.

\section{$N$ different hardcore interacting particles}\label{sec:N_particles}

For $N$ {\em identical } point particles we have the standard result
$\langle (x_{\cal T}-x_{{\cal T},0})^2 \rangle =(1/\varrho)
(4Dt/\pi)^{1/2}$ for the MSD of a tagged particle, where $\varrho=N/L$
is the concentration of particles ($N,L\rightarrow \infty$ with
$\varrho$ kept fixed). \cite{HA,LE,Lizana_Ambjornsson} For single-file
particles with {\em different} diffusion constants very recent results
show that the motion for a tagged particle stochastically jumping on a
lattice (exponential waiting time between jumps) is a fractional
Brownian motion (so that the PDF is Gaussian).
\cite{Jara_Gonzalves} The MSD was in Ref. \onlinecite{Jara_Gonzalves} shown
to take the same form as for identical diffusion constant but where
$D$ above is replaced by an effective diffusion constant, i.e., we
have:
  \be\label{eq:x2}
\langle (x_{\cal T}-x_{{\cal T},0} )^2\rangle = \kappa \left( \frac{4 D_{\rm eff} t}{\pi}\right)^{1/2}
  \ee
with the prefactor
  \be
\kappa=a \frac{1-f}{f}
  \ee
where $f=N/M$, $M$ is the number of lattice points and $a$ the lattice
spacing ($N,M\rightarrow \infty$ with $f$ fixed). In the continuum
limit (lattice spacing $a\rightarrow 0$) we get the point-particle
result $\kappa=1/\varrho$. In the continuum limit but with
finite-sized particles we have, as in Ref. \onlinecite{Lizana_Ambjornsson},
that $\kappa=(1-\varrho \Delta)/\varrho$, where $\Delta$ is the size
of the particles. The effective diffusion constant appearing in
Eq. (\ref{eq:x2}) is obtained by averaging the friction coefficients
(inverse of diffusion constants) according to:
  \be\label{eq:D_eff}
\frac{1}{D_{\rm eff}}={\rm lim}_{N\rightarrow \infty}\frac{1}{N}\sum_{i=1}^N \frac{1}{D_i}
  \ee
provided the limit on the right-hand side exists. \cite{Footnote1} The
result above is obtained for the (realistic) 'quenched' case, i.e.,
$D_i$ are kept the same for all the ensembles. The results above are
thus stronger than the results in Ref. \onlinecite{Aslangul_00} where
the 'annealed' case, i.e. for each ensemble the diffusion constants
are reshuffled between the particles, was studied. We also point out
that the results above are valid for an initial equilibrium density of
particles, whereas the results in Ref. \onlinecite{Aslangul_00} are limited to
the case that all particles are initially placed at the same point in
space.

\begin{figure}
 \includegraphics[width=8cm]{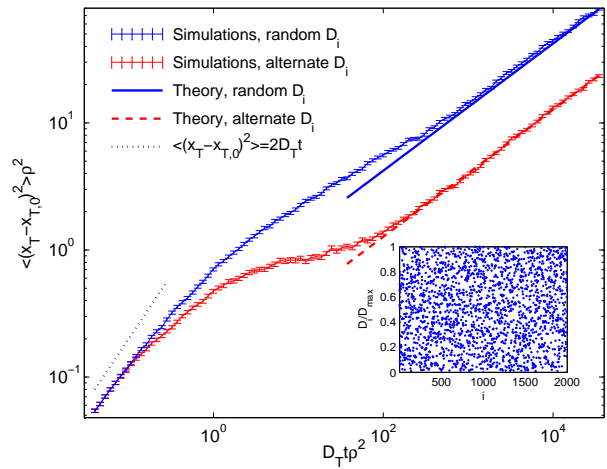}
 \caption{Mean square displacement for a tagged particle in the center
of a single-file system with different diffusion constants. The upper
(blue online) marks show the simulation results for a system where the
diffusion constants are 'quenched' and random with $0.01D_{\rm max}\le
D_i\le D_{\rm max}$ (the inset shows the diffusion constants
used). The lower (red online) marks are the results for an alternating
sequence of diffusion constants: $D_1=D_{\rm max}$,$D_2=0.01D_{\rm
max}$,$D_3=D_{\rm max}$, $D_4=0.01D_{\rm max}$ etc. For both cases the
tagged particle's diffusion constant $D_{\cal T}$ was set equal to
$D_{\rm max}$. The solid blue and dashed red line are the analytic
result as given in Eqs. (\ref{eq:x2})-(\ref{eq:D_eff}).  In all
simulations the tagged particle was initially placed at the center
lattice point and the non-tagged particles then randomly positioned to
the left and right of the tagged particle. The errorbars are the
standard errors. The following parameters were used: $M=10001$,
$N=2001$ and the number of ensembles $n_{\rm ens}=3200$. Without loss
of generality, we set $a=1$ and $D_{\cal T}=1$ (i.e., $a$ and $D_{\cal
T}$ determine the units of length and time in the problem) in all
simulations. Particle $1001$ was taken to be tagged. }
 \label{fig:msd}
\end{figure}
In Fig.  \ref{fig:msd} we show results of stochastic simulations for
$N=2001$ particles, with the middle particle (particle number $1001$)
being tagged.  The tagged particle is initially placed at the center
lattice point and the remaining particles are randomly positioned
(avoiding multiply occupied lattice sites \cite{Bebbington}) to the
left and right of the tagged particle for each ensemble. The details
of our stochastic scheme is presented in Appendix
\ref{sec:Gillespie}. Two cases are presented: the upper (blue)
marks shows simulations for the case of random 'quenched' distribution
of diffusion constants drawn between $0.01D_{\max}$ and $D_{\rm max}$.
The tagged particle diffusion constant $D_{\cal T}$ was set to $D_{\rm
max}$, and the diffusion constants used are shown in the inset in the
figure. The lower (red) marks represent results for an
alternating set of diffusion constants: the first particle has
diffusion constant $D_{\rm max}$, the second particles has $0.01D_{\rm
max}$ the third $D_{\rm max}$ etc. The tagged particle has diffusion
constant $D_{\rm max}$. For short times, $t\ll 1/(\varrho^2D_{\cal
T})$, we see that the case of 'quenched' random and alternating
diffusion constants give the same MSD (the tagged particle has a
diffusion constant equal to $D_{\rm max}$ in both cases). There has
been few collisions between particles and the MSD is proportional to
$t$, see Fig. \ref{fig:msd}. The MSD in the short-time regime is
slightly smaller than that of a free particle $\langle (x_{\cal
T}-x_{{\cal T},0} )^2\rangle= 2D_{\cal T} t$ [dotted line], simply due
to the fact that in some ensembles the tagged particle will initially
have a non-tagged particle at a neighbouring lattice site (every fifth
lattice site will on average contain a particle in the simulations in
the figure). For long times, $t\gg 1/(\varrho^2D_{\rm eff})$, there is
a cross-over to a single-file regime with the MSD proportional to
$t^{1/2}$. We notice an excellent agreement with the stochastic
simulations and the prediction in Eq. (\ref{eq:x2})-(\ref{eq:D_eff})
[solid blue and dashed red line] for long times. We point out that the
average diffusion constants [$(1/N)\sum_{i=1}^N D_i$ ] for the two
cases above are very close (more precisely, the two cases converge to
the same average diffusion constant for $N\rightarrow \infty$);
Fig. \ref{fig:msd} thus clearly illustrates that it is the average
friction coefficient which determine the long-time behavior for the
system rather than the average diffusion constant. For very long times
(beyond the time window in Fig. \ref{fig:msd}) and finite $L$, the
equilibrium PDF for the tagged particle should be reached, see
Ref. \onlinecite{Lizana_Ambjornsson} for an explicit expression.


\section{Summary and outlook} \label{sec:Summary}

In this study we have investigated the (single-file) dynamics of
hardcore interacting particles with different diffusion constants
diffusing in a one-dimensional system. For the two particle case we
obtained an analytic result for the conditional PDF (for arbitrary
initial particle positions and all times), from which we calculated
the tagged particle PDF, see Eq. (\ref{eq:P1}). For the general
$N$-particle case an asymptotic expression for the mean square
displacement of a tagged particle for long times was given,
Eq. (\ref{eq:x2}), and excellent agreement was found with our new
computationally efficient stochastic simulation technique based on the
Gillespie algorithm.

It will be interesting to see whether it is possible to generalize our
two-particle PDF to $N$ particles with different diffusion constants,
in order to access the full time behavior, and thus going beyond the
asymptotic results in Ref. \onlinecite{Jara_Gonzalves}. We point out
that the $N$ particle results given in this study assumed the mean
friction constant to be finite; we are currently considering the case
of a distribution of friction constants with diverging first moment.

The problem studied here, the dynamics of interacting species of
different kinds, shares many features with the dynamical behaviour of
cellular (and other biological) systems where heterogeneity and
interactions are important factors. We hope that our study will
inspire to further studies of many-body biology effects in living
systems.


\section{Acknowledgments}

We thank Milton Jara for sending an early version of
Ref. \onlinecite{Jara_Gonzalves} and for helpful correspondence. We are
grateful for discussions with Ophir Flomenbom. T.A. acknowledges the
support from the Knut and Alice Wallenberg Foundation. Part of this
research was supported by the NSF under grant CHE0556268, and the
Danish National Research Foundation via a grant to MEMPHYS. Computing
time was provided by the Danish Center for Scientific Computing at the
University of Southern Denmark.

\appendix

\section{New efficient stochastic algorithm based on the Gillespie algorithm}\label{sec:Gillespie}

\begin{figure}
 \includegraphics[width=8cm]{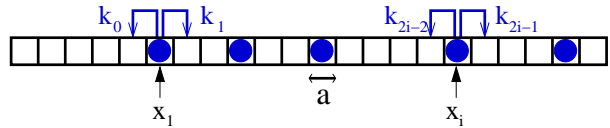}
 \caption{Schematic illustration of the lattice on which $N$ particles
   jump in the Gillespie simulation. The number of lattice sites are
   denoted by $M$ and the lattice spacing is $a$. Particle $i$ jumps to
   the left (right) with rate $k_{2i-2}$ ($k_{2i-1}$).  Note that
   these rates change during the simulation: if particle 1 is at the
   leftmost site then we impose the reflecting condition
   $k_1=0$. Similarly if particle $N$ is at the rightmost site we have
   $k_{2N-1}=0$. If particle $i$ and $i+1$ are at neighboring sites we set
   $k_{2i-1}=k_{2i}=0$ due to the hardcore repulsion. For the remaining
   configurations we have that the hop rates take the ``free''
   values: $k_\mu=k_\mu^f$. }
 \label{fig:Gillespie}
\end{figure}

Stochastic simulations using the Gillespie algorithm
\cite{DG1,DG2,DG3} is a convenient technique for
generating stochastic trajectories for interacting particles. Briefly,
we consider hopping of $N$ particles on a lattice with $M$ lattice
sites. The dynamics is governed by the `reaction' probability density (rPDF)
  \be\label{eq:RPDF}
P(\tau,\mu)=k_\mu \exp(- \sum_{\nu=0}^{2N-1} k_\nu \tau )
  \ee
where $\tau$ is the waiting time between jump events and $k_\mu$ are the
 jump rates. There are $2N$ jump rates for the process
considered here ($\mu=0,...,2N-1$): the rate for particle $1$ to $N$
jumping to the left and right respectively. We enumerate these rates
such that $k_{2i-2}$ is the rate for particle $i$ jumping to the left,
and $k_{2i-1}$ is the rate for particle $i$ jumping to the right, see
Fig. \ref{fig:Gillespie}. For the case that a particle have no
neighbors nor are at the end lattice points we set $k_\mu=k_\mu^f$
where $k_\mu^f$ are the ``free'' hop rates.
For the case that particle 1 (particle $N$) is at the leftmost
(rightmost) lattice point we have the reflecting condition $k_1=0$
($k_{2N-1}=0$). If two particles are at neighboring sites the hardcore
repulsion requires the right (left) rate for the leftmost (rightmost)
particle equals zero, i.e. if particle $i$ and $i+1$ are at
neighboring site we set $k_{2i-1}=k_{2i}=0$. Jump rates that are not
set to zero are equal to their ``free'' hop rates.  Thus, a stochastic
time series is generated through the steps: (1) place the particles at
their initial positions; (2) From the rPDF given in
Eq. (\ref{eq:RPDF}) we draw the random numbers $\tau$ (waiting time)
and $\mu$ (which particle to move and in what direction); (3) Update
the position of the chosen particle, the time $t$ and the rates $k_j$
for the new configuration and return to (2); (4) The loop (1)-(3) is
repeated until $t \ge t_{\rm stop}$, where $t_{\rm stop}$ is the stop
time for the simulation. This procedure produces a stochastic time
series for $0\le t\le t_{\rm stop}$. If steps (1)-(4) are repeated
$n_{\rm ens}$ times one obtains a histogram of particle positions (at
specified times). The ensemble averaged results of a Gillespie time
series is equivalent to the solution of a master equation
incorporating the rates given above; \cite{DG1,DG2} see for instance
Refs. \onlinecite{Golinelli_06} or \onlinecite{SC} for the explicit
expression for this master equation. In the limit $a\rightarrow 0$
with fixed diffusion constants $D_i=(k^f_{2i-2}+k^f_{2i-1})a^2/2$ the
master equation approaches the diffusion equation (assuming no drift,
i.e. that $k^f_{2i-2}=k^f_{2i-1}$) as specified in section
\ref{sec:ProblemDef} for an infinite system.

In the two subsequent subsections we consider two different methods
for generating a stochastic time series for the type of dynamics
described above: (A) the direct method and, and our new approach (B)
the trial-and-error method. We find that the latter method is superior
in computational speed whenever the fraction of occupied lattice sites
is not close to one.

\subsection{(A) The direct method}

Let us now review the ``direct method'' used for obtaining a
stochastic time series using the rPDF as
given in Eq. (\ref{eq:RPDF}). 
Briefly, the direct method as applied to the present type of system
involves the following steps \cite{DG1,DG2,DG3}
 \begin{enumerate}
\item Place the $N$ particles at their
initial positions and assign free hop rates $k_\mu^f$.
\item Draw two random numbers $r_1$ and $r_2$ ($0<r_1,r_2<1$)
\item The waiting time is obtained from the first random number as:
  \be
\tau=\frac{1}{\sum_{\mu=0}^{2N-1} k_\mu} \log \left( \frac{1}{r_1} \right)
  \ee 
\item The `reaction' $\mu$ is determined by using $r_2$ to determine
  the $\mu$ which satisfies
  \be
\sum_{\nu=0}^{\mu-1} k_\nu < r_2 \sum_{\nu=0}^{2N-1} k_\nu \le  \sum_{\nu=0}^{\mu} k_\nu
   \ee
\item Use the obtained $\mu$ and $\tau$ to update the particle
positions $X_i(t)$ and the time. After the chosen particle has been
moved, check the local environment for the particle and if necessary
update the rate constants $k_\mu$ accordingly ($k_\mu$ is either equal
to $k_\mu^f$ or 0).
\item Return to step (2).
\end{enumerate}
This procedure produces a stochastic time series $X_i(t)$ for the
particle positions. If steps (1)-(6) are repeated $n_{\rm ens}$ times
($n_{\rm ens}$ ensembles) one obtains a histogram of particle
positions (at specified times).

The direct method as applied to the present type of system is
computationally slow, since at each step the sum of all rate constants
has to be recalculated.

\subsection{ (B) The trial-and-error method}

In this subsection we present a new method for generating a time
series according to the rPDF as given in
Eq. (\ref{eq:RPDF}); we call this method the trial-and-error method.
\begin{enumerate}
\item Generate the partial sums of the {\em free} rate constants
  \bea\label{eq:p_mu}
p_0&=&0\nonumber\\
p_\mu&=&\sum_{\nu=0}^{\mu-1} k_\nu^f, \ \mu=1,...,2N
  \eea
\item Generate an initial configuration of particle positions.
\item Set the waiting time equal to zero, $\tau=0$.
\item Draw a random number $r_1$ ($0<r_1<1$). The waiting time is updated 
  according to:
  \be
\tau\rightarrow \tau+ \frac{1}{p_{2N}} \log \left( \frac{1}{r_1} \right)
  \ee 
\item Draw a random number $r_2$ ($0<r_2<1$). A trial `reaction' is
  determined by the $\mu$ which satisfies
  \be\label{eq:ineq}
p_\mu < r_2 p_{2N} \le  p_{\mu+1}
   \ee
\item Check if there is a reflecting wall or neighboring particle
  preventing the `reaction' $\mu$ from occurring, if so return to step
  (4). If $\mu$ is allowed then update the particle positions and
  the time.
\item Return to step (3).
  \end{enumerate}
The scheme above produces a stochastic time series $X_i(t)$, and if
steps (2)-(7) are repeated $n_{\rm ens}$ times ($n_{\rm ens}$
ensembles) one obtains a histogram of particle positions; note that
step (1) does not have to repeated for each ensemble. An efficient
method for performing the search for the $\mu$ satisfying the
inequality in Eq. (\ref{eq:ineq}) is presented in Appendix
\ref{sec:fast_method}. Notice that in the scheme above one does not
have to update the $k_\mu$ at each time step, since we are only using
the free rate constants $k_\mu^f$ (which are fixed at all times) in
the present scheme, nor do we need to perform a sum over all rate
constant at each time step. The only price we have to pay compared to
the direct method is the rejection step (6), which may cause us to
repeat steps (4) and (5) several times. However, whenever the fraction
of occupied lattice points is not close to one (or more precisely,
essentially whenever the occupation fraction is such that we can find
an allowed reaction in less than $N$ steps) we expect that the
trial-and-error method is faster than the direct method. A proof that
the direct and the trial-and-error methods are mathematically
equivalent is presented in Appendix \ref{sec:proof_two_methods}.

\section{Fast method for determining $\mu$ from Eq. (\ref{eq:ineq})}
\label{sec:fast_method}

In this appendix we give a simple, yet fast, algorithm for determining the
$\mu$ satisfying Eq. (\ref{eq:ineq}). We start by noticing that if all
$k_\mu^f$ are identical then we can directly satisfy
Eq. (\ref{eq:ineq}) by choosing $\mu=[2Nr_2]$, where $[z]$ gives the
largest integer smaller than $z$.  The algorithm below uses this fact
to directly find the correct $\mu$ for identical rate constants; for
non-identical rate constants we expect the number of attempts before the correct
$\mu$ is found to scale not worse than $\log N$ (see below).
The algorithm is:
  \begin{enumerate}
\item Initialize the two ``boundary'' parameters $\mu^{\rm left}$ and
  $\mu^{\rm right}$ to $\mu^{\rm left}=0$ and $\mu^{\rm
  right}=2N$. Also, introduce $p^{\rm left}$ and $p^{\rm right}$ which
  are initialized to the largest and smallest of the partial sums,
  i.e. we set initially $p^{\rm left}=p_0=0$ and $p^{\rm
    right}=p_{2N}$, see Eq. (\ref{eq:p_mu}).

\item Make a guess for $\mu$ using the formula
  \be\label{eq:mu_guess}
\mu_{\rm guess}=[\frac{r_2 p_{2N}-p^{\rm left}}{p^{\rm right}-p^{\rm left}} (\mu^{\rm right}-\mu^{\rm left})+\mu_{\rm left}].
  \ee
\item Check if the guess value for $\mu$ satisfies
  Eq. (\ref{eq:ineq}), if so terminate the loop.
\item If the guess value for $\mu$ {\em does not} satisfy
  Eq. (\ref{eq:ineq}) we separate between the cases: (a) if
  $r_2 p_{2N}\le p_{\mu_{\rm guess}}$ then move the right boundary parameters by
  changing $\mu^{\rm right}\rightarrow \mu_{\rm guess}$ and $p^{\rm
    right}\rightarrow p_{\mu_{\rm guess}}$; (b) if $r_2 p_{2N}>p_{\mu_{\rm
      guess}+1}$ then move the left boundary parameters by changing
  $\mu^{\rm left}\rightarrow \mu_{\rm guess}+1$ and $p^{\rm
    left}\rightarrow p_{\mu_{\rm guess}+1}$. Then return to step (2).
\end{enumerate}
The algorithm above will narrow down the search to a smaller and
smaller segment along the $\mu$-axis, until we manage to obtain the
correct $\mu$. The initial guess will always be $\mu_{\rm
  guess}=[2Nr_2]$, see Eq. (\ref{eq:mu_guess}), and hence for
identical free rate constants $k_0^f=k_1^f=...=k_{2N-1}^f$ the algorithm
above directly finds the correct $\mu$.
To argue for the maximally $\log N$-scaling for the number of iterations note
that if we had chosen $\mu_{\rm guess}=[(\mu^{\rm right}-\mu^{\rm left})/2]$ in
the algorithm then the number of steps to find the correct $\mu$ would be
maximally $[1+\log_2 N]$. The algorithm above should perform better than this
because the values of $\mu_{\rm guess}$ should approach the correct $\mu$
faster.

\section{Proof of equivalence of the direct method and the trial-and-error method}\label{sec:proof_two_methods}

Let us finally prove that indeed the trial-and-error method is
equivalent to the Gillespie rPDF,
Eq. (\ref{eq:RPDF}). More precisely we prove that the successful
`reaction' probability density function $P(\tau,\mu)$ agrees with
Eq. (\ref{eq:RPDF}).  Let us first define a trial-and-error relaxation
time as
  \be
\tau_{\rm TE}=\frac{1}{\sum_{\mu=0}^{2N-1} k_\mu^f}
  \ee
and we note that for each (successful or unsuccessful) attempt we draw a
 waiting time from the PDF:
  \be\label{eq:rho_TE}
\rho_{\rm TE}(\tau)=\frac{1}{\tau_{\rm TE}} e^{-\tau/\tau_{\rm TE}}
  \ee
see step (4) in the trial-and-error Gillespie scheme. To construct
$P(\tau,\mu)$ within the trial-and-error method we have to wait for a
{\em successful} attempt. We therefore separate between the following
(mutually exclusive) events:
  \begin{itemize}
\item success on the first attempt: the probability for this is 
  \be\label{eq:q}
\sigma_0=q=\frac{\sum_{\mu=0}^{2N-1} k_\mu}{\sum_{\mu=0}^{2N-1} k_\mu^f}=\frac{\tau_{\rm TE}}{\tau_G}
  \ee
where $\tau_G=1/[\sum_{\mu=0}^{2N-1} k_\mu]$ is the relaxation time for a successful event.
\item one failed attempt, then success: the probability for this is
  $\sigma_1=q(1-q)$.
\item .....
\item $m$ failed attempts and then success: the probability is
  $\sigma_m=q(1-q)^m$.
\item .....
  \end{itemize}
Given one of the above events the probability that reaction $\mu$ occurs is
\be
\rho_\mu=\frac{k_\mu}{\sum_{\nu=0}^{2N-1}k_\nu}=k_\mu\tau_G
\ee
The successful rPDF thus becomes
  \bea\label{eq:rho2}
P(\tau,\mu)&=&\rho_\mu\sigma_0 \rho_{\rm TE}(\tau) \nonumber\\
&& + \rho_\mu\sigma_1 \int_0^\tau d\tau_1 \rho_{\rm TE}(\tau_1) \rho_{\rm TE}(\tau-\tau_1) \nonumber\\
&&+\rho_\mu\sigma_2 \int d\tau_1 d\tau_2 \rho_{\rm TE}(\tau_1)\nonumber\\
&&\times \rho_{\rm TE}(\tau_2-\tau_1)\rho_{\rm TE}(\tau-\tau_2)+..
  \eea
The first term is the PDF for a successful move of type $\mu$ on the
first attempt. The second term is the PDF for a successful move of
type $\mu$ on the second attempt, and is obtained by considering
the waiting time density to first make an unsuccessful attempt at
time $\tau_1$ followed by a successful attempt at time $\tau$; since
$\tau_1$ can take any value between $0$ and $\tau$ we must integrate
over this interval. Similar arguments give expressions for higher
order terms.  In order to express the result given in
Eq.(\ref{eq:rho2}) in closed form we make a Laplace-transform which
gives:
  \bea\label{eq:P_mu_tau}
P(u,\mu)&=&\int_0^\infty du \ e^{-u\tau} P(\tau,\mu)= \rho_\mu\sigma_0 \rho_{\rm TE}(u)\nonumber\\
&&+\rho_\mu\sigma_1[\rho_{\rm TE}(u)]^2+\rho_\mu\sigma_2[\rho_{\rm TE}(u)]^3+..\nonumber\\
&=&\rho_\mu\sum_{n=1}^\infty \sigma_{n-1} [\rho_{\rm TE}(u)]^n\nonumber\\
&=&\rho_\mu q \rho_{\rm TE}(u) \sum_{n=0}^\infty [(1-q)\rho_{\rm TE}(u)]^n 
  \eea
where $\rho_{\rm TE}(u)=1/(1+u\tau_{\rm TE})$ is the Laplace-transform
of Eq. (\ref{eq:rho_TE}). Using the fact that the series in
Eq. (\ref{eq:P_mu_tau}) is a geometric series $\sum_{n=0}^\infty a^n
=1/(1-a)$, the explicit expression for $\rho_{\rm TE}(u)$ and
Eq. (\ref{eq:q}), we finally find:
  \be
P(u,\mu)=\frac{\rho_\mu}{1+u\tau_{\rm TE}/q}=\frac{k_\mu}{\tau_{\rm G}^{-1}+u}
  \ee
In the time domain this is the same as Eq. (\ref{eq:RPDF}), which
thereby completes the proof.




\begin{thebibliography}{10}
\bibitem{Footnote1}
The mathematical proof of Eqs. (\ref{eq:x2})-(\ref{eq:D_eff}) also requires the
  average friction for the particles to the left of the tagged particle should
  equal the average friction to the right, i.e. that ${\rm lim}_{N_L\rightarrow
  \infty} [ (1/N_L) \sum_{i \in {\rm left \ particles}} 1/D_i]= {\rm
  lim}_{N_R\rightarrow \infty} [(1/N_R) \sum_{i \in {\rm right \ particles}}
  1/D_i]$, where $N_L$ ($N_R$) is the number of particles to the left (right)
  of the tagged particle.
\end{thebibliography}
\end{document}